\documentclass[a4paper,twoside]{article}

\usepackage{epsfig}
\usepackage{subcaption}
\usepackage{calc}
\usepackage{amssymb}
\usepackage{amstext}
\usepackage{amsmath}
\usepackage{amsthm}
\usepackage{multicol}
\usepackage{pslatex}
\usepackage{float}
\usepackage{apalike}
\usepackage{SCITEPRESS}     

\begin{document}

\title{Modelling brain lesion volume in patches with CNN-based Poisson Regression}

\author{\authorname{Kevin Raina \sup{1}\orcidAuthor{0000-0002-6240-9675}}
\affiliation{\sup{1}Department of Mathematics and Statistics, University of Ottawa, Ontario, Canada}
\email{krain033@uottawa.ca} }


\abstract{Monitoring the progression of lesions is important for clinical response. Summary statistics such as lesion volume are objective and easy to interpret, which can help clinicians assess lesion growth or decay. CNNs are commonly used in medical image segmentation for their ability to produce useful features within large contexts and their associated efficient iterative patch-based training. Many CNN architectures require hundreds of thousands parameters to yield a good segmentation. In this work, an efficient, computationally inexpensive CNN is implemented to estimate the number of lesion voxels in a predefined patch size from magnetic resonance (MR) images. The output of the CNN is interpreted as the conditional Poisson parameter over the patch, allowing standard mini-batch gradient descent to be employed. The ISLES2015 (SISS) data is used to train and evaluate the model, which by estimating lesion volume from raw features, accurately identified the lesion image with the larger lesion volume for $86$\% of paired sample patches. An argument for the development and use of estimating lesion volumes to also aid in model selection for segmentation is made.}

\onecolumn \maketitle \normalsize \setcounter{footnote}{0} \vfill

\section{\uppercase{Introduction}}
\label{sec:introduction}

\noindent Many segmentation challenges have been undertaken recently, showing the need for automated models in clinical settings \cite{maier2017isles,winzeck2018isles,bakas2018identifying}. Along with strong predictive power, these challenges stress the importance of fast inference, as lesions can quickly spread. For instance, ischemic stroke lesions cause increasing tissue death within hours of onset, requiring reperfusion therapies around this time. The stroke progresses through  acute, sub-acute and chronic stages within days. Gliomas, the most common type of malignant brain tumour, grows at a rate of increasing severity depending on the grade. As the tumour gets larger, symptoms often worsen, reinforcing the need to monitor lesion growth. 

A simple, yet powerful, summary statistic is the lesion volume, or when the brain is represented as labelled voxels, the lesion label count. In a clinical study, \cite{alexander2010correlating} showed that lesion volume is a significant covariate for understanding ischemic stroke deficits after the initial onset. In application, lesion volume has generally been a dependable factor in the prognosis of ischemic stroke \cite{merino2007lesion,rivers2006acute} and multiple sclerosis \cite{zivadinov2012abnormal,bagnato2011lesions}. The objectivity of counts enables straightforward inference: a higher lesion label count means the lesion has grown.

In comparison to segmenting entire 3D medical images, directly estimating the number of lesion voxels in the brain from raw features should be expected to require less computational resources since it is no longer necessary to provide detailed information about the lesion's appearance. Nonetheless, in a study by \cite{erskine2005resolution} comparing the effects on volume estimation by using different magnetic resonance imaging scanners, lesion volume was estimated from a computer-assisted segmentation tool. Other methods for estimating lesion volume favored a geometric approach, wherein the lesion's surface area per slice is calculated and the estimate is derived by summing accross slices \cite{park2013semi,filippi1995resolution}. In contrast, the output of our proposed statistical direct lesion counting model is a single non-negative integer that doesn't require sophisticated viewing software or significant memory usage. 

CNNs have shown promising results in lesion segmentation, as in the works of \cite{kamnitsas2017efficient}, \cite{havaei2017brain} and \cite{ronneberger2015u}, due to their ability to produce useful features from 

\begin{figure*}
	\centering
	\includegraphics[scale=0.8]{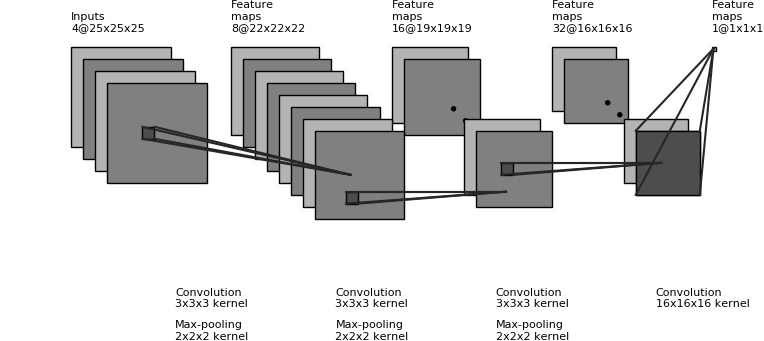}
	\caption{3D architecture employed for counting lesions. The input tensor is obtained by stacking patches from the patient's brain MRI over $4$ different modalities. After applying convolution and pooling operations, the final output is a real number.} \label{fig:PRCountNet}
\end{figure*}

\noindent large visual spatial contexts combined with efficient iterative patch-based training and dense inference. The output of the CNN is often interpeted as the parameter of a conditional distribution. For instance, in \cite{kamnitsas2017efficient,havaei2017brain}, the output at each voxel is the parameter of a Bernoulli conditional distribution. The Poisson distribution is generally well known for modelling counts over time and space, and particularly has been applied to modelling the count of multiple sclerosis lesions over time \cite{altman2005application,albert1994time}. For this reason, we propose the lesion label counts, or equivalently lesion volume, in a predefined patch size is assumed to follow a Poisson distribution conditional on the patch features. The CNNs of \cite{kamnitsas2017efficient} and \cite{havaei2017brain} use hundreds of thousands of parameters for segmentation. Using CNNs, coupled with good distributional assumptions, should allow for smaller architectures and faster convergence on the counting task. 

One prior related work by \cite{dubost2017gp} used a 3D CNN, similar to U-Net \cite{ronneberger2015u} to predict global lesion label count, but produces a segmentation when testing by using a removable global pooling layer. A drawback of estimating global lesion label counts is the need for more patient brain images, since an entire brain serves as a single sample. In their study, training was performed on $1,289$ 3D PD-weighted MRI scans, whereas some challenges provide limited training instances \cite{maier2017isles}. Another challenge with global lesion counts is being able to efficiently produce scalar outputs from larger 3D information, which would require additional preprocessing and transformations. Estimation of counts in patches can help in these situations.                     

This paper is organized as follows: Section 2 describes the methods, Section 3 presents the results, and a discussion follows in Section 4. 

\section{\uppercase{Methods}}

\subsection{Architecture}
\noindent The proposed network is shown in Figure \ref{fig:PRCountNet}. As input it stacks $25 \times 25 \times 25$ patches from each MR sequence, runs 3 layers of convolution and max pooling which have sizes $3 \times 3 \times 3$ and $2 \times 2 \times 2$ respectively, followed by a final convolution of size $16 \times 16 \times 16$ to output one real number. In addition to the convolution and max pooling operations, Leaky ReLU nonlinearity was used. It is important to note the number of output activations at each layer are considerably small to reduce the total number of parameters. The patch is not only a parameter of the training features, but is intertwined with the task as well, since it delineates the region over which the lesion label count is estimated.

\begin{figure*}
	\centering
	\includegraphics[scale=0.6]{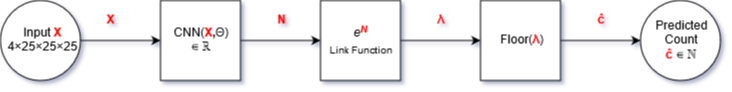}
	\caption{Block diagram representation of the CNN-based Poisson Regression model. The predicted count is obtained by flooring the estimated conditional Poisson parameter ($\lambda$).} \label{fig:Block}
\end{figure*}

\paragraph{Training}: A block diagram of the methodology is shown in Figure \ref{fig:Block}. In accordance with the notation of Figure \ref{fig:Block}, the architecture associates one real number $N$ to each input tensor. Then, the model can be formulated as: $c | X \sim Pois(e^{N(X,\Theta)})$, where $c$ is the lesion label count over the patch, X are the input features used in the architecture, and $\Theta$ are the parameters of the 

\noindent architecture. In order to train the parameters from observed counts $c_i$ (assumed to Poisson distributed with rate $\lambda_i$), mini-batch gradient descent with a batch size $b$ is used to minimize the average negative log-likelihood, $-\frac{1}{b}\sum_{i=1}^{b}log(\frac{\lambda_i^{c_i}e^{-\lambda_i}}{c_i!})$, plus additional L1 and L2 regularization terms to prevent overfitting. The samples used in the mini-batch were taken so as to ensure the lesion count was non-zero by insisting the central voxel be lesion. Since training only samples non-zero counts, it will not be efficient at predicting counts for completely randomly sampled patches, for which zero counts are more frequent. A possible workaround for this task is using a zero-inflated Poisson model (ZIP) \cite{lambert1992zero}, which is suggested for a future study.

\subsection{Implementation Details}

\noindent The open-source software Tensorflow was used to implement the model \cite{abadi2016tensorflow}. Non-zero counts were sampled in mini-batches of size $10$. Weights were initialized under a Gaussian with mean $0$ and standard deviation of $0.001$, while biases are initialized to $0$. Moreover, the Adam Optimizer was used with initial learning rate of $10^{-4}$, and training was stopped when the average cost over $1,000$ iterations increased. This always happened within $15,000$ iterations. In comparison, the segmentation CNN of \cite{kamnitsas2017efficient} has default training configurations set to $70,000$ iterations, demonstrating a quick ability to learn for direct counting models. L1 regularization and L2 regularization were used and set to $10^{-8}$ and $10^{-6}$ respectively. In addition to regularization, dropout on all hidden layers was employed at a rate ot $0.5$. At prediction time, the mean Poisson rate was floored to provide an integer estimate. Table 1 summarizes the implementation details.

\begin{table}[H]
	\centering
	\caption{Numerical summary of implementation details.}
	\label{tab1}
	\begin{tabular}{|c|c|}
		\hline
		\bfseries Implementation Detail  &  \bfseries Value\\ 
		\hline
		Batch size ($b$) & $10$\\
		\hline
		Kernel initialization mean(std.) & $0(0.001)$ \\
		\hline
		Learning rate& $10^{-4}$\\
		\hline
		L1, L2 coefficients & $10^{-8}, 10^{-6}$ \\
		\hline
		Dropout & $0.5$\\
		\hline
		\end{tabular}
\end{table}

\section{Experiments and Results}

\subsection{Dataset}

\noindent The architecture and model were trained and evaluated on the ISLES2015 (SISS) training data, which consists of $28$ patients with sub-acute ischemic stroke. All data come from the same clinical center, which provided $4$ MR sequences for each patient: FLAIR, DWI, T1 and T1-contrast. Images are of size $230 \times 230 \times 153/154$, are processed to not contain the skull and have isotropic $1mm^3$ voxel resolution. Sub-acute ischemic stroke lesions have large variation in size. For instance, in the dataset, the smallest lesion consists of $106$ voxels, and the largest consists of $233,547$ voxels. From the $28$ brains, $20$ were randomly selected to form the training set, and the remaining $8$ formed the validation set. This selection was carried out once, and the same split was used across all experiments. It should be acknowledged that the choice of the training and validation split will have an effect on performance due to the aformentioned variability in lesion count.    

\subsection{Model Performance}

\begin{figure}[H]
	\centering
	\includegraphics[scale=0.35]{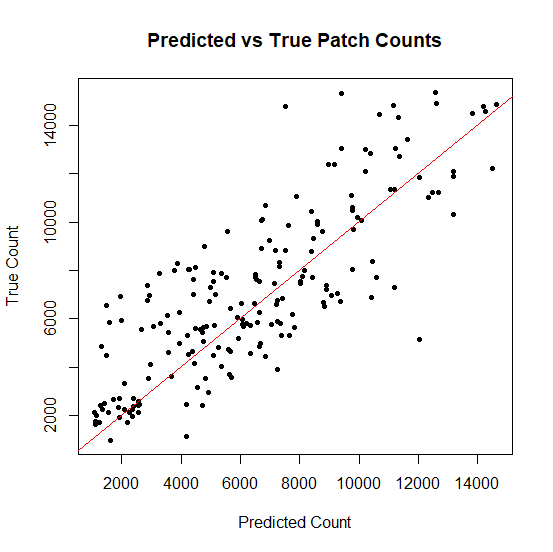}
	\caption{Plot of true count and estimated count for 200 lesion patch samples. Coefficient of multiple correlation (Pearson's correlation coefficient between predicted and actual values) of $R = 0.81$.} \label{fig:RPlot}
\end{figure}

\noindent To evaluate the performance of the architecture, several metrics are calculated on $10,000$ patches sampled to have a non-zero lesion label count from the validation set. Sampling was done by first selecting the validation brain and then selecting a patch from the brain. Estimated counts that surpassed the possible count in the predefined patch size were adjusted to predict the maximum possible count. In the experiment applied to the ISLES2015 (SISS) data, the mean absolute error (MAE) rounded up to the nearest integer was computed to be $1,458$. In addition, over the same samples the average estimated count to true count ratio was computed to be $1.15$. Finally, the mean relative error (MRE) was $0.42$ for the ischemic stroke lesions. True patch lesion label counts vary from a few hundred to $15,000$, indicating a promising initial result. Figure \ref{fig:RPlot} plots the estimated and true counts for 200 samples.

\subsection{Predicting count order}

\begin{figure}[H]
	\centering
	\includegraphics[scale=0.76]{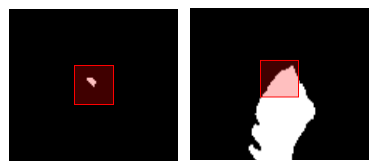}
	\caption{Example of predicting count order, where the red square outline represents the middle slice of the $25^3$ patch. The true counts from left to right are: $356$ and $5297$. The predicted counts from left to right are: $501$ and $5640$. In this case, the model accurately identifies the left image as having the smaller lesion volume.  }
	\label{pair}
\end{figure}

\noindent The second experiment was to order pairs of patches by lesion label count, which can be applied in a clinical setting to compare lesion images over time and assess growth or decay. Given any two image patches containing lesions, the goal is to evaluate the model's ability to identify the image with the larger (equivalently smaller) lesion volume using the proposed estimation. In the experiment, $10,000$ pairs of non-zero count patches are sampled from the validation fold. A sample is counted as correct if the predicted counts preserve the same order as the true counts. Running the experiment on the ISLES2015 (SISS) data gives a correct order prediction for $86$ \% of samples, demonstrating good ordering capabilities. Figure \ref{pair} shows an accurately predicted sample.

\section{\uppercase{Future Work}}

\subsection{Extension to Arbitrary Patches}

\noindent The analysis and modelling undertaken in the previous sections were done on patches that contain lesion voxels. That is, the patches were sampled to contain lesion voxels. Although it was shown that this has the potential for modelling lesion growth or decay by first estimating lesion volume, relaxing this restriction allows for the prediction of counts in arbitrary patches, which are more frequently zero counts. Due to the unbalanced nature of the data, one proposition for a future study is to combine a zero-inflated Poisson model which is known to account for excess zero counts in data, with CNNs.  

\subsection{Location Detection}

Being able to predict counts in arbitrary patches can form the basis for 
lesion location detection. A possible algorithm could be to randomly sample patches from one brain, predict their counts, and record the central voxel 
position for the patch with the maximum predicted count. The voxel position returned by the
algorithm should identify an area of significant lesion presence in the brain, 
assuming the model is well-tuned. Though simple, the algorithm is theoretically
stable when applied on a single lesion, since applying true counts in place of predictions will always return
the location of a lesion provided a lesion is present.

Rather than recording just the maximum predicted count from the given sample,
another option would be to order the predicted counts for samples drawn from one brain. From a pre-defined quantile, larger counts a long with their central voxel positions will delineate the lesion. 

\subsection{Model Selection for Segmentation}

The size of a lesion is one of the factors that can allow some configurations of a segmentation algorithm to perform better than others \cite{kamnitsas2017efficient}. Smaller lesions, often found in sub-acute ischemic stroke, are usually tougher to segment and produce lower dice coefficients \cite{havaei2017brain} than larger lesions due to a relatively higher number of false positives compared to true positives. For CNNs, configurations mainly pertain to the architecture employed. Given that some architectures may segment smaller lesions better, one option is to regress the architecture employed for a brain, on an estimate of the global lesion label count in the brain.  

\section{\uppercase{Discussion}}

Direct counting models might be useful in clinical settings. Some potential examples are but not limited to: monitoring and locating lesions. They can also aid segmentation by selecting configurations of segmentation algorithms based on global lesion size, from raw data, and reducing the required segmentation area through lesion detection. Further developments need to be made to predict patch counts, including: improving accuracy in the predictions of non-zero counts, accounting for highly imbalanced zero counts, developing sampling-based algorithms for lesion location detection, and providing aggregate patch measures to predict global lesion count. This will increase the effectiveness and broaden the applications of direct counting models.

\bibliographystyle{apalike}
{\small
\bibliography{Raina2020BIOIMAGING2}}

\end{document}